\documentclass{ifacconf}%
\usepackage{graphicx}
\usepackage{natbib}
\usepackage{amssymb}
\usepackage{amsmath}
\usepackage{amsfonts}
\usepackage{multirow}
\usepackage{epsfig}%
\begin{document}
\begin{frontmatter}
\title{Multi-intersection Traffic Light Control Using Infinitesimal Perturbation
Analysis\thanksref{footnoteinfo}}
\thanks[footnoteinfo]{The authors' work is supported in part by NSF under
Grant EFRI-0735974, by AFOSR under grant FA9550-09-1-0095, by DOE under grant
DE-FG52-06NA27490, and by ONR under grant N00014-09-1-1051.}
\author[First]{Yanfeng Geng}
\author[First]{Christos G. Cassandras}
\address[First]{Division of Systems Engineering and Center for Information and Systems Engineering, 	Boston University, Brookline, MA, 02446 (e-mail: gengyf@bu.edu, cgc@bu.edu)}
\begin{abstract}                
We address the traffic light control problem for multiple intersections in tandem by
viewing it as a stochastic hybrid system and developing a Stochastic Flow
Model (SFM) for it. Using Infinitesimal Perturbation Analysis (IPA), we derive
on-line gradient estimates of a cost metric with respect to the controllable
green and red cycle lengths. The IPA estimators obtained require counting
traffic light switchings and estimating car flow rates only when specific
events occur. The estimators are used to iteratively adjust light cycle
lengths to improve performance and, in conjunction with a standard
gradient-based algorithm, to obtain optimal values which adapt to changing
traffic conditions. Simulation results are included to illustrate the approach.
\end{abstract}
\end{frontmatter}

\section{Introduction}

The Traffic Light Control (TLC) problem aims at dynamically controlling the
flow of traffic at an intersection through the timing of green/red light
cycles with the objective of reducing congestion, hence also the delays
incurred by drivers. The more general problem involves a set of intersections
and traffic lights with the objective of reducing overall congestion over an
area covering multiple urban blocks. Since the control of one intersection
influences the traffic flow from or towards others, this is a complex problem
further complicated by the fact that traffic flows constantly change depending
on the time of day, accidents, weather conditions, etc. Recent technological
developments involving better, inexpensive sensors and wireless sensor
networks have enabled the collection of data (e.g., counting vehicles in a
specific road section) which can be used to optimally select traffic light
cycles over specific time intervals in a day or even to dynamically control
them based on real-time data. Thus, methodologies that would not be possible
to implement not long ago are now becoming feasible. The approach proposed in
this paper to the TLC problem is specifically intended to exploit these recent developments.

Several different approaches have been proposed to solve the TLC problem. It
is formulated as a Mixed Integer Linear Programming (MILP) problem in
\cite{Dujardin11}, and as an Extended Linear Complementary Problem (ELCP) in
\cite{Schutter99}. A Markov Decision Process (MDP) approach has been proposed
in \cite{Yu06} and Reinforcement Learning (RL) was used in \cite{Thorpe97},
with several extensions found in \cite{Prashanth11, Wiering04}. A game
theoretic viewpoint is given in \cite{Alvarez10}, while a hybrid system
formulation is presented in \cite{Zhao03}. Due to its complexity when viewed
as an optimization problem, fuzzy logic is often used in both a single
(isolated) junction \cite{Murat05} and multiple junctions \cite{Choi02}.
Expert systems \cite{Findler92} and evolutionary algorithms \cite{Taale98}
have also been applied to develop a traffic light controller for a single
intersection. Perturbation analysis techniques were used in \cite{Head96} and
a formal approach using Infinitesimal Perturbation Analysis (IPA) to solve the
TLC problem was presented in \cite{Panayiotou05} for a single intersection.

In \cite{GengCDC12}, we study the TLC problem for a single intersection using
a Stochastic Flow Model (SFM) and Infinitesimal Perturbation Analysis (IPA).
In this paper, we extend our analysis to two tandem intersections. We still
adopt a stochastic hybrid system modeling framework (see \cite{Cassandras08}%
,\cite{Cassandras06}), since the problem involves both event-driven dynamics
in the switching of traffic lights and time-driven dynamics that capture the
flow of vehicles through an intersection. Although one can also view this as a
purely Discrete Event System (DES) with the intersection area as a
\textquotedblleft server\textquotedblright\ processing \textquotedblleft
users\textquotedblright\ (vehicles), the fact that a vehicle does not
exclusively occupy this area makes a flow-based viewpoint a more accurate way
to model such a process. While in most traditional flow models the flow rates
involved are treated as deterministic parameters, a SFM as introduced in
\cite{Cassandras02} treats them as stochastic processes. In the TLC problem,
this is consistent with continuously and randomly varying traffic flows,
especially in heavy traffic conditions where the problem is most interesting.
With only minor technical assumptions imposed on the properties of such
processes, a general IPA theory for stochastic hybrid systems was recently
presented in \cite{Wardi10},\cite{Cassandras10} through which one can estimate
on line gradients of certain performance measures with respect to various
controllable parameters. These estimates may be incorporated in standard
gradient-based algorithms to optimize system parameter settings. IPA was
originally developed as a technique for evaluating gradients of sample
performance functions in queueing systems and using them as unbiased gradient
estimates of performance metrics. However, IPA estimates become biased when
dealing with aspects of queueing systems such as multiple user classes,
blocking due to limited resource capacities, and various forms of feedback
control. The use of IPA in stochastic hybrid systems, however, circumvents
these limitations and yields simple unbiased gradient estimates (under mild
technical conditions) of useful metrics even in the presence of blocking and a
variety of feedback control mechanisms (see \cite{Yao11}.)

In Section 2, we formulate the TLC problem for two intersections and construct
a SFM. In Section 3, we derive an IPA estimator for a cost function gradient
with respect to a controllable parameter vector defined by green and red cycle
lengths. This is then used to iteratively adjust these cycle lengths to
improve performance and, under proper conditions, obtain optimal parameter
values. Simulation-based examples are given in Section 4 and we conclude with
Section 5.

\section{Problem Formulation}

\begin{figure}[tbh]
\centering
\includegraphics[scale = 0.4]{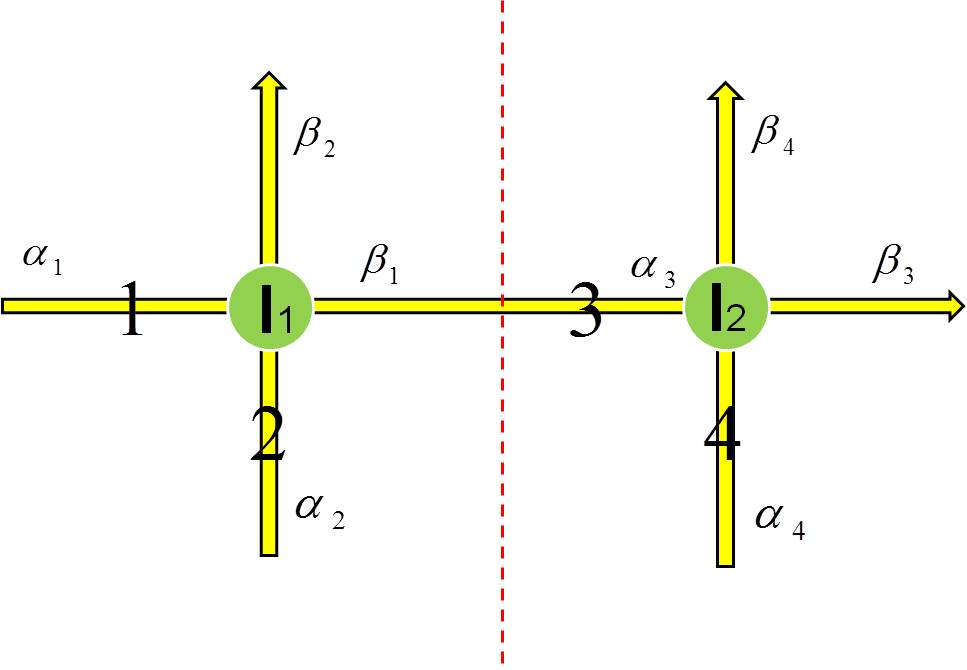} \caption{Two tandem
traffic intersections}%
\label{junction}%
\end{figure}

In this paper, we concentrate on solving the TLC problem for two coupled
intersections, as shown in Fig. \ref{junction}. There are four roads and four
traffic lights, with each traffic light controlling the\ associated incoming
traffic flow. The traffic in road 1 of intersection $I_{1}$ flows into road 3
of $I_{2}$. For simplicity, we make the following assumptions: $(i)$ Left-turn
and right-turn traffic flows are not considered, i.e., traffic lights only
control vehicles going straight. $(ii)$ A YELLOW light is combined with a RED
light (therefore, the YELLOW light duration is not explicitly controlled).
$(iii)$ Road 3 is long enough that cars accumulated in it do not influence the
departure process of road 1.

The system involves a number of stochastic processes which are all defined on
a common probability space $(\Omega,F,P)$. Each of the four roads is
considered as a queue with a random \emph{arrival} flow process $\{\alpha
_{n}(t)\},n=1,\ldots,4.$, where $\alpha_{n}(t)$ is the instantaneous vehicle
arrival rate at time $t$. When the traffic light corresponding to road $n$ is
GREEN, the \emph{departure} flow process is denoted by $\{\beta_{n}%
(t)\},n=1,\ldots,4.$ Let the GREEN light duration in a cycle of queue $n$ be
$\theta_{n}$, and the controllable parameter vector of interest is
$\theta=[\theta_{1},\ldots,\theta_{4}]$. We define a state vector
$x(\theta,t)=[x_{1}(\theta,t),...,x_{4}(\theta,t)]$ where $x_{n}(\theta
,t)\in\mathbb{R}^{+}$ is the content of queue $n=1,...,4$. We use the notation
$x_{n}(\theta,t)$ to emphasize the dependence of the queue content on $\theta
$; however, for notational simplicity, we will write $x_{n}(t)$ when no
confusion arises. We also define a left-continuous \textquotedblleft
clock\textquotedblright\ state variable $z_{n}(t)$, $n=1,...,4$, associated
with the GREEN light cycle for queue $n$ as follows:
\begin{align}
\dot{z}_{n}(t) &  =\left\{
\begin{array}
[c]{ll}%
1 & \text{if }0<z_{n}(t)<\theta_{n}\text{ or }z_{\bar{n}}(t)=\theta_{\bar{n}%
}\\
0 & \text{otherwise}%
\end{array}
\right.  \label{dzdt}\\
z_{n}(t^{+}) &  =0\text{ if }z_{n}(t)=\theta_{n}\nonumber
\end{align}
where $\bar{n}$ is the index of the road perpendicular to road $n$ at the same
intersection. We set $z(t)=[z_{1}(t),...,z_{4}(t)]$. Thus, $z_{n}(t)$ measures
the time since the last switch from RED to GREEN of the traffic light for
queue $n$. It is reset to $0$ as soon as the GREEN cycle length $\theta_{n}$
is reached and remains at this value while the light is GREEN for queue
$\bar{n}$; as soon as that cycle ends, i.e., $z_{\bar{n}}(t)=\theta_{\bar{n}}%
$, then $\dot{z}_{n}(t)=1$ and the process repeats.

To simplify notation, we set $B_{n}(z,\theta)=1$ if the Boolean expression
used in (\ref{dzdt}), i.e., $[0<z_{n}(t)<\theta_{n}$ or $z_{\bar{n}}%
(t)=\theta_{\bar{n}}]$, is true (light is GREEN) and $B_{n}(z,\theta)=0$
otherwise. We can now write the dynamics of each state variable $x_{n}(t)$ as
follows:%
\begin{equation}
\dot{x}_{n}(t)=\left\{
\begin{array}
[c]{ll}%
\alpha_{n}(t) & \text{if }B_{n}(z,\theta)=0\\
0 & \text{if }x_{n}(t)=0\text{ and }\alpha_{n}(t)\leq\beta_{n}(t)\\
\alpha_{n}(t)-\beta_{n}(t) & \text{otherwise}%
\end{array}
\right.  \label{dxdt}%
\end{equation}
where
\begin{equation}
\beta_{n}(t)=\left\{
\begin{array}
[c]{ll}%
h_{n}(t) & \text{if }B_{n}(z,\theta)=1\text{ and }x_{n}(t)>0\\
\alpha_{n}(t) & \text{if }B_{n}(z,\theta)=1\text{ and }x_{n}(t)=0\\
0 & \text{otherwise}%
\end{array}
\right.  \label{departRate}%
\end{equation}
In (\ref{departRate}), $h_{n}(t)$ describes the departure process if the road
is not empty. According to assumption $(iii)$, $\beta_{1}(t)$ is independent
of $x_{3}(t)$. However,  $\alpha_{3}(t)$ depends on the departure process
$\beta_{1}(t)$ of queue 1; in particular,
\begin{equation}
\alpha_{3}(t)=\beta_{1}(t)\label{alpha3}%
\end{equation}
Combining (\ref{dxdt}) through (\ref{alpha3}), we have the dynamics of queue
3:
\begin{equation}
\dot{x}_{3}(t)=\left\{
\begin{array}
[c]{ll}%
h_{1}(t) & \text{if }B_{3}(z,\theta)=0,\text{ }B_{1}(z,\theta)=1,\\
& \text{ and }x_{1}(t)>0\text{ \ \ \ \ \ \ \ \ \ \ \ \ \ \ \ \ \ }%
(5.1)\\
\alpha_{1}(t) & \text{if }B_{3}(z,\theta)=0,\text{ }B_{1}(z,\theta)=1,\\
& \text{ and }x_{1}(t)=0\text{ \ \ \ \ \ \ \ \ \ \ \ \ \ \ \ \ \ }%
(5.2)\\
0 & \text{if }B_{3}(z,\theta)=0,\text{ }B_{1}(z,\theta)=0,\\
& \text{ or }B_{3}(z,\theta)=1,\text{ }x_{3}(t)=0\text{ \ }(5.3)\\
h_{1}(t)-h_{3}(t) & \text{if }B_{3}(z,\theta)=1,\text{ }B_{1}(z,\theta)=1,\\
& \text{ and }x_{3}(t)>0,\text{ }x_{1}(t)>0\text{ \ \ \ }(5.4)\\
\alpha_{1}(t)-h_{3}(t) & \text{if }B_{3}(z,\theta)=1,\text{ }B_{1}%
(z,\theta)=1,\\
& \text{ and }x_{3}(t)>0,\text{ }x_{1}(t)=0\text{ \ \ \ }(5.5)\\
-h_{3}(t) & \text{if }B_{3}(z,\theta)=1,\text{ }B_{1}(z,\theta)=0,\\
& \text{ and }x_{3}(t)>0\text{ \ \ \ \ \ \ \ \ \ \ \ \ \ \ \ \ \ }(5.6)\\
&
\end{array}
\right.  \label{f3}%
\end{equation}

The operation of the intersection can be viewed as a hybrid system with the
time-driven dynamics described by (\ref{dxdt})-(\ref{f3}) and event-driven
dynamics dictated by GREEN-RED light switches and by events causing some
$x_{n}(t)$ to switch from positive to zero or vice versa.

\begin{figure}[tbh]
\centering
\includegraphics[scale = 0.30]{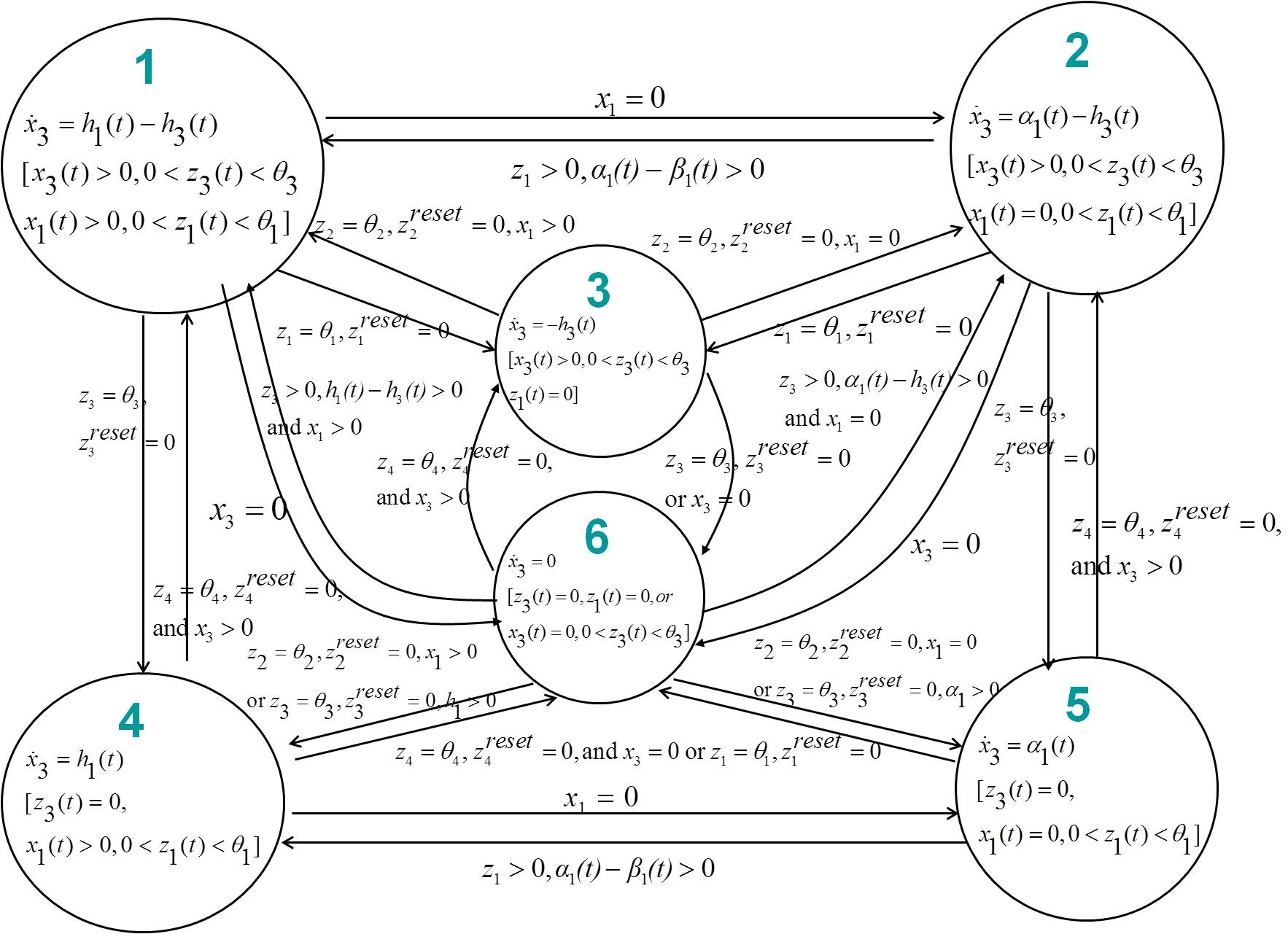} \caption{The Stochastic Hybrid
Automaton model}%
\label{Automaton}%
\end{figure}

\begin{figure}[tbh]
\centering
\includegraphics[scale = 0.4]{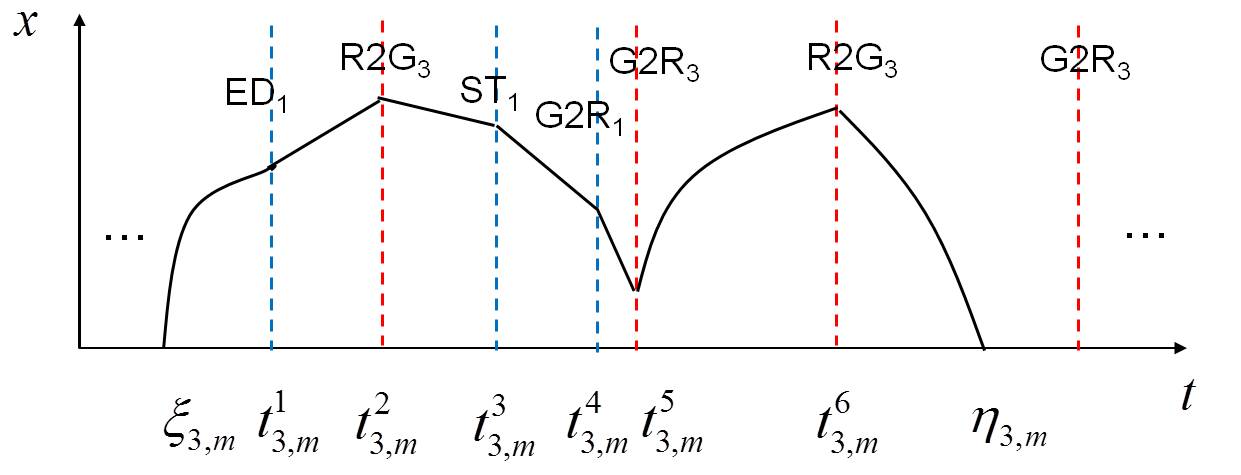}\caption{A typical sample path of
traffic light queue 3}%
\label{SamplePath}%
\end{figure}

Using the standard definition of a Stochastic Hybrid Automaton (SHA) (e.g.,
see \cite{Cassandras08}), we may obtain a SHA model for queue 1,2 and 4 which
is similar to \cite{GengCDC12}. Here, we concentrate on the SHA for the
operation of queue 3 as shown in Fig. \ref{Automaton}. This reflects the fact
that a typical sample path of any one of the queue contents (as shown in Fig.
\ref{SamplePath}) consists of intervals over which $x_{n}(t)>0$, which we call
Non-Empty Periods (NEPs), followed by intervals where $x_{n}(t)=0$, which we
call Empty Periods (EPs). Thus, the entire sample path consists of a series of
alternating NEPs and EPs. The event set that
affects any queue $n=1,2,4$ is $\Phi_{n}=\{e_{1},e_{2},e_{3},e_{4},e_{5}\}$
where $e_{1}$ is a switch in the sign of $\alpha_{n}(t)-\beta_{n}(t)$ from
non-positive to strictly positive, $e_{2}$ is a switch in the sign of
$\alpha_{n}(t)$ from $0$ to strictly positive, $e_{3}$ is the queue content
becoming empty, i.e., $x_{i}=0$, which terminates a NEP (and initiates an EP),
$e_{4}$ switches a light from RED to GREEN, and $e_{5}$ switches a light from
GREEN to RED. For easier reference, we label $e_{3}$ as \textquotedblleft%
$E_{n}$\textquotedblright\ for the end of NEP events, $e_{4}$ as
\textquotedblleft$R2G_{n}$\textquotedblright\ and $e_{5}$ as \textquotedblleft%
$G2R_{n}$\textquotedblright\ for the light switching events. The resulting
start of a NEP is an event \textquotedblleft induced\textquotedblright\ by
either $e_{5}$ or $e_{2}$ or $e_{1}$ which we will refer to as an
\textquotedblleft$S_{n}$\textquotedblright\ event. For queue 3, the event set
includes all those events that cause a jump in the value of $\dot{x}_{3}(t)$
in (\ref{f3}). As we can see from Fig. \ref{Automaton}, every event of
$\Phi_{1}$ also affects the dynamics of queue 3. Thus, we have $\Phi
_{3}=\{S_{1},E_{1},R2G_{1},G2R_{1},S_{3},E_{3},R2G_{3},G2R_{3}\}$.

Returning to Fig. \ref{SamplePath}, the $m$th NEP in a sample path of queue 3,
$m=1,2,\ldots$, is denoted by $[\xi_{3,m},\eta_{3,m})$, i.e., $\xi_{3,m}$,
$\eta_{3,m}$ are the occurrence times of the $m$th $S_{3}$ and $E_{3}$ event
respectively at this queue. During the $m$th NEP, $t_{3,m}^{j}$,
$j=1,\ldots,J_{3,m}$, denotes the time when an event occurs.

\begin{figure}[tbh]
\centering
\includegraphics[scale = 0.6]{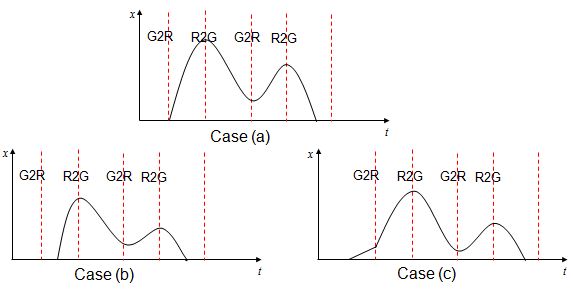} \caption{Three ways for starting
a NEP}%
\label{StartofNEP}%
\end{figure}

Our objective is to select $\theta$ so as to minimize a cost function that
measures a weighted mean of the queue lengths over a fixed time interval
$[0,T]$. In particular, we define the sample function%
\begin{equation}
L(\theta;x(0),z(0),T)=\frac{1}{T}\sum\limits_{n=1}^{4}\int\nolimits_{0}%
^{T}w_{n}x_{n}(\theta,t)dt\label{costfun1}%
\end{equation}
where $w_{n}$ is a cost weight associated with queue $n$ and $x(0),z(0)$ are
given initial conditions. It is obvious that since $x_{n}(t)=0$ during EPs of
queue $n$, we can rewrite (\ref{costfun1}) in the form
\begin{equation}
L(\theta;x(0),z(0),T)=\frac{1}{T}\sum\limits_{n=1}^{4}\sum\limits_{m=1}%
^{M_{n}}\int\nolimits_{\xi_{n,m}}^{\eta_{n,m}}w_{n}x_{n}(\theta
,t)dt\label{costfun}%
\end{equation}
where $M_{n}$ is the total number of NEPs during the sample path of queue $n$.
For convenience, we also define%
\begin{equation}
L_{n,m}(\theta)=\int\nolimits_{\xi_{n,m}}^{\eta_{n,m}}x_{n}(\theta
,t)dt\label{Lnm}%
\end{equation}
to be the sample cost associated with the $m$th NEP of queue $n$. We can now
define our overall performance metric as%
\begin{equation}
J(\theta;x(0),z(0),T)=E\left[  L(\theta;x(0),z(0),T\right]  \label{costfun3}%
\end{equation}
Since we do not impose any limitations on the processes $\{\alpha_{n}(t)\}$
and $\{\beta_{n}(t)\}$, it is infeasible to obtain a closed-form expression of
$J(\theta;x(0),z(0),T)$. The only assumption we make is that $\alpha_{n}(t)$,
$\beta_{n}(t)$ are piecewise continuous w.p. 1. The value of IPA as developed
for general stochastic hybrid systems in \cite{Cassandras10} is in providing
the means to estimate the performance metric gradient $\nabla J(\theta)$, by
evaluating the sample gradient $\nabla L(\theta)$. As shown elsewhere (e.g.,
see \cite{Cassandras10}), these estimates are unbiased under mild technical
conditions. Moreover, an important property of IPA estimates is that they are
often independent of the unknown processes $\{\alpha_{n}(t)\}$ and
$\{\beta_{n}(t)\}$ or they depend on values of $\alpha_{n}(t)$ or $\beta
_{n}(t)$ at specific event times only. Such robustness properties of IPA
(formally established in \cite{YaoCgc11}) make it attractive for estimating on
line performance sensitivities with respect to controllable parameters such as
$\theta$ in our case. One can then use this information to either improve
performance or, under appropriate conditions, solve an optimization problem
and determine an optimal $\theta^{\ast}$ through an iterative scheme:
\begin{equation}
\theta_{i,k+1}=\theta_{i,k}-\gamma_{k}H_{i,k}(\theta_{k},x(0),T,\omega
_{k}),k=0,1,...\label{iteration}%
\end{equation}
where $H_{i,k}(\theta_{k},x(0),T,\omega_{k})$ is an estimate of $dJ/d\theta
_{i}$ based on the information obtained from the sample path denoted by
$\omega_{k}$, and $\gamma_{k}$ is the stepsize at the $k$th iteration. Next we
will focus on how to obtain $dL/d\theta_{i}$, $i=1,2,3,4$. We may then also
obtain $\theta^{\ast}$ through (\ref{iteration}), provided that the random
processes $\{\alpha_{n}(t)\}$ and $\{\beta_{n}(t)\}$ are stationary over
$[0,T]$. We will assume that the derivatives $dL/d\theta_{i}$, exist for all
$\theta_{i}\in\mathbb{R}^{+}$ w.p. 1 (if this is violated, IPA is still
possible by considering one-sided derivatives; see \cite{Cassandras02}.)

\section{Infinitesimal Perturbation Analysis (IPA)}

Consider a sample path of the system as modeled in Fig. \ref{Automaton} over
$[0,T]$ and let $\tau_{k}(\theta)$ denote the occurrence time of the $k$th
event (of any type), where we stress its dependence on $\theta$. To simplify
notation, we define the derivatives of the states $x_{n}(t,\theta)$ and
$z_{n}(t,\theta)$ and event times $\tau_{k}(\theta)$ with respect to
$\theta_{i}$, $i=1,\ldots,4$, as follows:
\begin{equation}
x_{n,i}^{\prime}(t)\equiv\frac{\partial x_{n}(\theta,t)}{\partial\theta_{i}%
},\text{ }z_{n,i}^{\prime}(t,\theta)\equiv\frac{\partial z_{n}(\theta
,t)}{\partial\theta_{i}},\text{ }\tau_{k,i}^{\prime}\equiv\frac{\partial
\tau_{k}(\theta)}{\partial\theta_{i}}\label{IPAnotation}%
\end{equation}
Taking derivatives with respect to $\theta_{i}$ in (\ref{costfun}), we obtain%
\begin{align*}
\frac{dL(\theta)}{d\theta_{i}} &  =\frac{1}{T}\sum\limits_{n=1}^{4}%
\sum\limits_{m=1}^{M_{n}}\left[  \int\nolimits_{\xi_{n,m}}^{\eta_{n,m}}%
w_{n}x_{n,i}^{\prime}(t)dt\right.  \\
&  +w_{n}x_{n}(\eta_{n,m})\frac{\partial\eta_{n,m}}{\partial\theta_{i}%
}-\left.  w_{n}x_{n}(\xi_{n,m})\frac{\partial\xi_{n,m}}{\partial\theta_{i}%
}\right]
\end{align*}
Since, at the start and end of a NEP $x_{n}(\xi_{n,m})=x_{n}(\eta_{n,m})=0$,
this reduces to%
\begin{align}
\frac{dL(\theta)}{d\theta_{i}} &  =\frac{1}{T}\sum\limits_{n=1}^{4}%
\sum\limits_{m=1}^{M_{n}}\int\nolimits_{\xi_{n,m}}^{\eta_{n,m}}w_{n}%
x_{n,i}^{\prime}(t)dt\label{dLi}\\
&  \equiv\frac{1}{T}\sum\limits_{n=1}^{4}\sum\limits_{m=1}^{M_{n}}w_{n}%
\frac{dL_{n,m}(\theta)}{d\theta_{i}}\nonumber
\end{align}
where the last equality follows from the definition (\ref{Lnm}).

By assumption $(iii)$, $\beta_{1}(t)$ is independent of $x_{3}(t)$. Therefore,
$dL_{n,m}/d\theta_{i}=0$ for $n=1,2$ and $i=3,4$. It follows that
$x_{n,i}^{\prime}(t)$ for $n=1,2$ and $i=1,2$ can be obtained by the analysis
of a single isolated intersection in \cite{GengCDC12}. Since equation
(\ref{dxdt}) can still be applied for $x_{4}(t)$, we can obtain $x_{4,i}%
^{\prime}(t)$, $i=1,\ldots,4$, similar to a queue in an isolated intersection.
Therefore, in what follows, we focus on obtaining $x_{3,i}^{\prime}(t)$ and
hence $dL_{3,m}/d\theta_{i}$.

\subsection{State Derivatives}

Observe that the determination of the sample derivatives in (\ref{dLi})
depends on the state derivatives $x_{n,i}^{\prime}(t)$. The purpose of IPA is
to evaluate these derivatives as functions of observable sample path
quantities. We pursue this next, using the framework established in
\cite{Cassandras10} where, for arbitrary stochastic hybrid systems, it is
shown that the state and event time derivatives in (\ref{IPAnotation}) can be
obtained from three fundamental \textquotedblleft IPA
equations\textquotedblright. For the sake of self-sufficiency, these equations
are rederived here as they pertain to our specific SFM. Looking at
(\ref{dxdt}) and Fig. \ref{Automaton}, note that the dynamics of $x_{n}(t)$
are fixed over any interevent interval $[\tau_{k},\tau_{k+1})$ and we write
$\dot{x}_{n}(t)=f_{n,k}(t)$ to represent the appropriate expression on the
right-hand-side of (\ref{dxdt}) over this interval. We have
\[
x_{n}(t)=x_{n}(\tau_{k})+\int\nolimits_{\tau_{k}}^{t}f_{n,k}(\tau)d\tau
\]
and taking derivatives with respect to $\theta_{i}$, we get
\begin{align}
&  x_{n,i}^{\prime}(t)=x_{n,i}^{\prime}(\tau_{k}^{-})+\frac{\partial
x_{n}(\tau_{k}^{-})}{\partial t}\tau_{k,i}^{\prime}\label{xprime}\\
&  +\int\nolimits_{\tau_{k}}^{t}\left[  \frac{\partial f_{n,k}(\tau)}{\partial
x_{n}}x_{n,i}^{\prime}(\tau)+\frac{\partial f_{n,k}(\tau)}{\partial\theta_{i}%
}\right]  d\tau-f_{n,k}(\tau_{k})\tau_{k,i}^{\prime}\nonumber
\end{align}
Letting $t=\tau_{k}^{+}$ and since $\frac{\partial x_{n}(\tau_{k}^{-}%
)}{\partial t}=f_{n,k-1}(\tau_{k}^{-})$ and $\frac{\partial f_{n,k}}{\partial
x_{n}}=\frac{\partial f_{n,k}}{\partial\theta_{i}}=0$ from (\ref{dxdt}), we
obtain
\begin{equation}
x_{n,i}^{\prime}(\tau_{k}^{+})=x_{n,i}^{\prime}(\tau_{k}^{-})+[f_{n,k-1}%
(\tau_{k}^{-})-f_{n,k}(\tau_{k}^{+})]\tau_{k,i}^{\prime}\label{jumps}%
\end{equation}
Moreover, taking derivatives with respect to $t$ in (\ref{xprime}), we get,
for all $t\in\lbrack\tau_{k},\tau_{k+1})$,
\begin{equation}
\frac{d}{dt}x_{n,i}^{\prime}(t)=\frac{\partial f_{n,k}}{\partial x_{n}%
}(t)x_{n,i}^{\prime}(t)+\frac{\partial f_{n,k}}{\partial\theta_{i}%
}(t)\label{dxdtdtheta}%
\end{equation}
Again, $\frac{\partial f_{n,k}}{\partial x_{n}}=\frac{\partial f_{n,k}%
}{\partial\theta_{i}}=0$ and we get $\frac{d}{dt}x_{n,i}^{\prime}(t)=0$.
Therefore, $x_{n,i}^{^{\prime}}(t)$ remains constant over all $t\in\lbrack
\tau_{k},\tau_{k+1})$:
\begin{equation}
x_{n,i}^{\prime}(t)=x_{n,i}^{\prime}(\tau_{k}^{+}),\text{ \ \ \ }t\in
\lbrack\tau_{k},\tau_{k+1})\label{xprime(t)}%
\end{equation}
Thus, focusing on a NEP of $x_{n}(t)$, the queue content derivative is
piecewise constant with jumps occurring according to (\ref{jumps}). The final
step is to obtain the event time derivatives $\tau_{k,i}^{\prime}$ appearing
in (\ref{jumps}), which we do next.

\subsection{Event Time Derivatives}

Clearly $\tau_{k,i}^{\prime}$ depends on the type of event occurring at time
$\tau_{k}$. Following the framework in \cite{Cassandras10}, there are three
types of events for a general stochastic hybrid system. For the purpose of
these definitions, let the continuous state component of the hybrid system be
$\mathbf{x}\in X\subseteq\mathbb{R}^{N}$, $\mathbf{x=[}x_{1},\ldots x_{N}]$,
and let $\theta\in\Theta\subseteq\mathbb{R}^{M}$.

\begin{enumerate}
\item \textbf{Exogenous Events.} An event is \emph{exogenous} if it causes a
discrete state transition at time $\tau_{k}$ independent of the controllable
parameter $\theta$. Thus, it satisfies
\begin{equation}
\tau_{k,i}^{\prime}=0
\end{equation}

\item \textbf{Endogenous Events.} An event is occurring at time $\tau_{k}$ is
\emph{endogenous} if there exists a continuously differentiable function
$g_{k}:\mathbb{R}^{N}\times\Theta\rightarrow\mathbb{R}$ such that
\begin{equation}
\tau_{k,i}=\min\{t>\tau_{k-1}:g_{k}(x(\theta,t),\theta)=0\}
\end{equation}
where the function $g_{k}$ normally corresponds to a guard condition in a
hybrid automaton. Taking derivatives with respect to $\theta_{i}$,
$i=1,\ldots,m$, it is straightforward to obtain
\begin{equation}
\tau_{k,i}^{\prime}=-\frac{\frac{\partial g_{k}}{\partial\theta_{i}}%
+\sum_{j=1}^{N}\frac{\partial g_{k}}{\partial x_{j}}x_{j,i}^{\prime}(\tau
_{k}^{-})}{\sum_{j=1}^{N}\frac{\partial g_{k}}{\partial x_{j}}f_{j,k-1}%
(\tau_{k}^{-})} \label{dtaudtheta}%
\end{equation}

\item \textbf{Induced Events.} An event at time $\tau_{k}$ is \emph{induced}
if it is triggered by the occurrence of another event at time $\tau_{m}%
\leq\tau_{k}$. In this case, $\tau_{k}^{\prime}$ depends on the derivative
$\tau_{m}^{\prime}$ (details can be found in \cite{Cassandras10}.)
\end{enumerate}

In the following, we consider each of the event types at queue $3$ that were
identified in the previous section and derive the corresponding event time
derivatives. Based on these, we can then also derive the state derivatives
through (\ref{jumps}) and (\ref{xprime(t)}).

\textbf{(1)} \emph{Event }$E_{1}$\emph{ ends a NEP of queue 1}. This is an
endogenous event that occurs when $x_{1}(\theta,t)=0$. Thus, when such an
event occurs at $\tau_{k}$, let $g_{k}(x(\theta,t),\theta)=x_{1}(\theta,t)=0$.
Using (\ref{dtaudtheta}), we get $\tau_{k,i}^{\prime}=\frac{-x_{1,i}^{\prime
}(\tau_{k}^{-})}{f_{1,k-1}(\tau_{k}^{-})}$. Looking at (\ref{f3}), we have
either $f_{3,k-1}(\tau_{k}^{-})=h_{1}(\tau_{k}^{-})-h_{3}(\tau_{k}^{+})$ and
$f_{3,k}(\tau_{k}^{+})=\alpha_{1}(\tau_{k}^{-})-h_{3}(\tau_{k}^{+})$ when
$B_{3}(z,\theta)=1$, or $f_{3,k-1}(\tau_{k}^{-})=h_{1}(\tau_{k}^{-})$ and
$f_{3,k}(\tau_{k}^{+})=\alpha_{1}(\tau_{k}^{-})$ when $B_{3}(z,\theta)=0$. In
both cases, $f_{3,k-1}(\tau_{k}^{-})-f_{3,k}(\tau_{k}^{+})=h_{1}(\tau_{k}%
^{-})-\alpha_{1}(\tau_{k}^{-})$. Using these values in (\ref{jumps}) along
with $\tau_{k,i}^{\prime}$ above we get%
\begin{align}
x_{3,i}^{\prime}(\tau_{k}^{+}) &  =x_{3,i}^{\prime}(\tau_{k}^{-})-\frac
{[h_{1}(\tau_{k}^{-})-\alpha_{1}(\tau_{k}^{-})]x_{1,i}^{\prime}(\tau_{k}^{-}%
)}{\alpha_{n}(\tau_{k}^{-})-h_{1}(\tau_{k}^{-})}\nonumber\\
&  =x_{3,i}^{\prime}(\tau_{k}^{-})+x_{1,i}^{\prime}(\tau_{k}^{-}),\quad
i=1,\ldots,4\label{type1}%
\end{align}
As we can see, $x_{3,i}^{\prime}(\tau_{k}^{+})$ explicitly depends on
$x_{1,i}^{\prime}(\tau_{k}^{-})$.

\textbf{(2)} \emph{Event }$E_{3}$\emph{ ends a NEP of queue 3}. This is an
endogenous event that occurs when $x_{3}(\theta,t)=0$. Thus, when such an
event occurs at $\tau_{k}$, let $g_{k}(x(\theta,t),\theta)=x_{3}(\theta,t)=0$.
Using (\ref{dtaudtheta}), we get $\tau_{k,i}^{\prime}=\frac{-x_{3,i}^{\prime
}(\tau_{k}^{-})}{f_{3,k-1}(\tau_{k}^{-})}$. According to (\ref{f3}), we have
$f_{3,k}(\tau_{k}^{+})=0$. Using these values in (\ref{jumps}) along with
$\tau_{k,i}^{\prime}$ above we get%
\[
x_{3,i}^{\prime}(\tau_{k}^{+})=x_{3,i}^{\prime}(\tau_{k}^{-})-(f_{3,k-1}%
(\tau_{k}^{-})-0)\frac{x_{3,i}^{^{\prime}}(\tau_{k}^{-})}{f_{3,k-1}(\tau
_{k}^{-})}=0
\]
Thus, at the end of a NEP $[\xi_{3,m},\eta_{3,m})$ of queue $3$ we have%
\begin{equation}
x_{3,i}^{\prime}(\eta_{3,m}^{+})=0,\quad i=1,\ldots,4\label{E_event}%
\end{equation}
indicating that these state derivatives are always reset to $0$ upon ending a NEP.

\textbf{(3)} \emph{Event }$G2R_{1}$\emph{, i.e., the GREEN light of queue 1
switches to RED}. This is an endogenous event that occurs when $g_{k}%
(x(\theta,t),\theta)=z_{1}(\tau_{k})=\theta_{1}$. $\tau_{k,i}^{\prime}$ is
determined by the following lemma.

\begin{lem}
Let $\zeta_{1,k}$ be the total number of $G2R_1$ events that have occurred before or at $\tau_{k}$, and $\rho_{1,k}$ be the total number of $R2G_1$ events that have occurred before or at $\tau_{k}$. Then, $\tau_{k,1}^{^{\prime }}= \zeta_{1,k}$, $\tau_{k,2}^{^{\prime }}= \rho_{1,k}$, $\tau_{k,3}^{^{\prime }}=0$ and $\tau_{k,4}^{^{\prime }}=0$
\label{lemma1}
\end{lem}

The proof of this lemma can be found in \cite{GengCDC12}. According to
(\ref{f3}), we have either $f_{3,k-1}(\tau_{k}^{-})-f_{3,k}(\tau_{k}%
^{+})=h_{1}(\tau_{k}^{-})$ (from (5.1)-(5.3), or (5.4)-(5.6)), or
$f_{3,k-1}(\tau_{k}^{-})-f_{3,k}(\tau_{k}^{+})=\alpha_{1}(\tau_{k}^{-})$ (from
(5.2)-(5.3), or (5.5)-(5.6)). From (\ref{departRate}), we can combine these
two situations and simply so that $f_{3,k-1}(\tau_{k}^{-})-f_{3,k}(\tau
_{k}^{+})=\beta_{1}(\tau_{k}^{-})$. According to (\ref{jumps}), we get
\begin{equation}
x_{3,i}^{^{\prime}}(\tau_{k}^{+})=\left\{
\begin{array}
[c]{ll}%
x_{3,i}^{^{\prime}}(\tau_{k}^{-})+\beta_{1}(\tau_{k}^{-})\zeta_{1,k} & i=1\\
x_{3,i}^{^{\prime}}(\tau_{k}^{-})+\beta_{1}(\tau_{k}^{-})\rho_{1,k} & i=2\\
x_{3,i}^{^{\prime}}(\tau_{k}^{-}) & i=3,4
\end{array}
\right.  \label{type3di}%
\end{equation}

\textbf{(4)} \emph{Event }$G2R_{3}$\emph{, i.e., the GREEN light of queue 3
switches to RED}. This is an endogenous event that occurs when $g_{k}%
(x(\theta,t),\theta)=z_{3}(\tau_{k})=\theta_{3}$. $\tau_{k,i}^{\prime}$ is
determined by the following lemma.

\begin{lem}
Let $\zeta_{3,k}$ be the total number of $G2R_3$ events that have occurred before or at $\tau_{k}$, and $\rho_{3,k}$ be the total number of $R2G_3$ events that have occurred before or at $\tau_{k}$. Then, $\tau_{k,3}^{^{\prime }}= \zeta_{3,k}$, $\tau_{k,4}^{^{\prime }}= \rho_{4,k}$, $\tau_{k,1}^{^{\prime }}=0$ and $\tau_{k,2}^{^{\prime }}=0$
\label{lemma1}
\end{lem}

From (\ref{f3}), if $x_{3}(\tau_{k}^{-})>0$, we have $f_{3,k-1}(\tau_{k}%
^{-})-f_{3,k}(\tau_{k}^{+})=-h_{3}(\tau_{k}^{-})$ (from (5.4)-(5.1), or
(5.5)-(5.2), or (5.6)-(5.3)). According to (\ref{jumps}), the state derivative
is
\begin{equation}
x_{3,i}^{^{\prime}}(\tau_{k}^{+})=\left\{
\begin{array}
[c]{ll}%
x_{3,i}^{^{\prime}}(\tau_{k}^{-}) & i=1,2\\
x_{3,i}^{^{\prime}}(\tau_{k}^{-})-h_{3}(\tau_{k}^{-})\zeta_{3,k} & i=3\\
x_{3,i}^{^{\prime}}(\tau_{k}^{-})-h_{3}(\tau_{k}^{-})\rho_{3,k} & i=4
\end{array}
\right.  \label{type4di1}%
\end{equation}
If $x_{3}(\tau_{k}^{-})=0$, $f_{3,k-1}(\tau_{k}^{-})-f_{3,k}(\tau_{k}%
^{+})=-\beta_{1}(\tau_{k}^{+})$ (from (5.3)-(5.1), or (5.3)-(5.2)). Then,
\begin{equation}
x_{3,i}^{^{\prime}}(\tau_{k}^{+})=\left\{
\begin{array}
[c]{ll}%
x_{3,i}^{^{\prime}}(\tau_{k}^{-}) & i=1,2\\
x_{3,i}^{^{\prime}}(\tau_{k}^{-})-\beta_{1}(\tau_{k}^{+})\zeta_{3,k} & i=3\\
x_{3,i}^{^{\prime}}(\tau_{k}^{-})-\beta_{1}(\tau_{k}^{+})\rho_{3,k} & i=4
\end{array}
\right.  \label{type4di2}%
\end{equation}

\textbf{(5)} \emph{Event }$R2G_{1}$\emph{, i.e., the RED light of queue 1
switches to GREEN}. This is an endogenous event that occurs when
$g_{k}(x(\theta,t),\theta)=z_{2}(\tau_{k})=\theta_{2}$. $\tau_{k,i}^{\prime}$
is determined by Lemma 1. Similar to the analysis of a $G2R_{1}$ event, we
have $f_{3,k-1}(\tau_{k}^{-})-f_{3,k}(\tau_{k}^{+})=-\beta_{1}(\tau_{k}^{+})$
(from (5.3)-(5.1), or (5.6)-(5.4), or (5.3)-(5.2), or (5.6)-(5.5)). Thus the
state derivative is
\begin{equation}
x_{3,i}^{^{\prime}}(\tau_{k}^{+})=\left\{
\begin{array}
[c]{ll}%
x_{3,i}^{^{\prime}}(\tau_{k}^{-})-\beta_{1}(\tau_{k}^{+})\zeta_{1,k} & i=1\\
x_{3,i}^{^{\prime}}(\tau_{k}^{-})-\beta_{1}(\tau_{k}^{+})\rho_{1,k} & i=2\\
x_{3,i}^{^{\prime}}(\tau_{k}^{-}) & i=3,4
\end{array}
\right.  \label{type5di}%
\end{equation}

\textbf{(6)}\emph{Event }$R2G_{3}$\emph{, i.e., the RED light of queue 3
switches to GREEN}. This is an endogenous event that occurs when
$g_{k}(x(\theta,t),\theta)=z_{4}(\tau_{k})=\theta_{4}$. $\tau_{k,i}^{\prime}$
is determined by Lemma 2. From (\ref{f3}), we have $f_{3,k-1}(\tau_{k}%
^{-})-f_{3,k}(\tau_{k}^{+})=h_{3}(\tau_{k}^{+})$ (from (5.1)-(5.4), or
(5.2)-(5.5), or (5.3)-(5.6)). The state derivative is
\begin{equation}
x_{3,i}^{^{\prime}}(\tau_{k}^{+})=\left\{
\begin{array}
[c]{ll}%
x_{3,i}^{^{\prime}}(\tau_{k}^{-}) & i=1,2\\
x_{3,i}^{^{\prime}}(\tau_{k}^{-})+h_{3}(\tau_{k}^{+})\zeta_{3,k} & i=3\\
x_{3,i}^{^{\prime}}(\tau_{k}^{-})+h_{3}(\tau_{k}^{+})\rho_{3,k} & i=4
\end{array}
\right.  \label{type6di}%
\end{equation}

\textbf{(7)} \emph{Event }$S_{1}$\emph{ starts a NEP of queue 1} As already
mentioned, this is an event induced by a $G2R_{1}$ event ($e_{5}$) or a switch
of $\alpha_{1}(t)$ from zero to a strictly positive value ($e_{2}$) occurring
during a RED cycle, or a switch of $\alpha_{1}(t)-\beta_{1}(t)$ from a
non-positive to a strictly positive value ($e_{1}$) occurring during a GREEN
cycle (see Fig. \ref{StartofNEP}). Consequently, there are three possible
cases to consider as follows.

\textbf{Case (7a)}: \emph{A NEP of queue 1 starts right after a }$G2R_{1}%
$\emph{ event}. This is an endogenous event and was analyzed in Case
\textbf{(3)}. Since $x_{1}(\tau_{k}^{-})=0$, $f_{3,k-1}(\tau_{k}^{-}%
)-f_{3,k}(\tau_{k}^{+})=\alpha_{1}(\tau_{k}^{-})$ (from (5.2)-(5.3) or from
(5.5)-(5.6) in (\ref{f3})). We get
\begin{equation}
x_{3,i}^{^{\prime}}(\tau_{k}^{+})=\left\{
\begin{array}
[c]{ll}%
x_{3,i}^{^{\prime}}(\tau_{k}^{-})+\alpha_{1}(\tau_{k}^{-})\zeta_{1,k} & i=1\\
x_{3,i}^{^{\prime}}(\tau_{k}^{-})+\alpha_{1}(\tau_{k}^{-})\rho_{1,k} & i=2\\
x_{3,i}^{^{\prime}}(\tau_{k}^{-}) & i=3,4
\end{array}
\right.  \label{type7adi}%
\end{equation}

\textbf{Case (7b)}: \emph{ A NEP of queue 1 starts while }$z_{1}(\tau_{k}%
)=0$\emph{,  }$z_{2}(\tau_{k})>0$. This is an exogenous event occurring during
a RED cycle for queue $1$ and is due to a change in $\alpha_{1}(\tau_{k})$
from a zero to a strictly positive value. Therefore, $\tau_{k,i}^{\prime}=0$.
We then have
\begin{equation}
x_{3,i}^{^{\prime}}(\tau_{k}^{+})=x_{3,i}^{^{\prime}}(\tau_{k}^{-}%
),i=1,2,3,4\label{7b}%
\end{equation}

\textbf{Case (7c)}: \emph{A NEP of queue 1 starts while }$z_{2}(\tau_{k}%
)=0$\emph{,  }$z_{1}(\tau_{k})>0$. This is an exogenous event occurring during
a GREEN cycle for queue $1$ due to a change in $\alpha_{1}(\tau_{k})$ or
$\beta_{1}(\tau_{k})$ that results in $\alpha_{1}(\tau_{k})-\beta_{1}(\tau
_{k})$ switching from a non-positive to a strictly positive value. The
analysis is exactly the same as Case (\textbf{7b}) above and (\ref{7b}) applies.

\textbf{(8)} \emph{Event }$S_{3}$\emph{ starts a NEP of queue 3}. This is
similar to Case \textbf{(7)}, and there are also three possible cases to consider.

\textbf{Case (8a)}: \emph{A NEP of queue 3 starts right after a }$G2R_{3}%
$\emph{ event}. This is an endogenous event and was analyzed in Case
\textbf{(4)}. Since $x_{3}(\tau_{k}^{-})=0$, we have $f_{3,k-1}(\tau_{k}%
^{-})-f_{3,k}(\tau_{k}^{+})=-\beta_{1}(\tau_{k}^{+})$, and (\ref{type4di2})
applies. Suppose that this is the $m$th NEP, i.e., $\tau_{k}=\xi_{3,m}$. We
have already shown in (\ref{E_event}) that $x_{3,i}^{\prime}(\eta_{n,m-1}%
^{+})=0$. In addition, we have $x_{3}(t)=0$ over the interval $[\eta
_{3,m-1},\xi_{3,m})$, thus $x_{3,i}^{\prime}(t)=0$ for all $t\in\lbrack
\eta_{3,m-1},\xi_{3,m})$ and we get $x_{3,i}^{^{\prime}}(\tau_{k}^{-}%
)=x_{3,i}^{^{\prime}}(\xi_{k}^{-})=0$ , the state derivative in
(\ref{type4di2}) becomes
\begin{equation}
x_{3,i}^{^{\prime}}(\tau_{k}^{+})=\left\{
\begin{array}
[c]{ll}%
0 & i=1,2\\
-\beta_{1}(\tau_{k}^{+})\zeta_{3,k} & i=3\\
-\beta_{1}(\tau_{k}^{+})\rho_{3,k} & i=4
\end{array}
\right.  \label{type8adi}%
\end{equation}

\textbf{Case (8b)}: \emph{ A NEP of queue 3 starts while }$z_{3}(\tau_{k}%
)=0$\emph{, \ }$z_{4}(\tau_{k})>0$. This is due to a change in $\alpha
_{3}(\tau_{k})$ from a zero to a strictly positive value. It also happens in
two ways. First, $\alpha_{3}(\tau_{k})=\beta_{1}(\tau_{k})$ becomes positive
because a $G2R_{1}$ event occurs. Then (\ref{type3di}) applies, where
$x_{3,i}^{^{\prime}}(\tau_{k}^{-})=0$. Second, $\beta_{1}(\tau_{k})$ becomes
positive because either $h_{1}(\tau_{k})$ or $\alpha_{1}(\tau_{k})$ switches
from 0 to a strictly positive value, which is an exogenous event. Therefore,
the state derivative is
\begin{equation}
x_{3,i}^{^{\prime}}(\tau_{k}^{+})=x_{3,i}^{^{\prime}}(\tau_{k}^{-}%
)=0,i=1,2,3,4\label{8b}%
\end{equation}

\textbf{Case (8c)}: \emph{A NEP of queue 3 starts while }$z_{4}(\tau_{k}%
)=0$\emph{, \ }$z_{3}(\tau_{k})>0$. This is due to a change in $\alpha
_{3}(\tau_{k})-\beta_{3}(\tau_{k})$ from a zero to a strictly positive value,
which may happen in two ways. First, it becomes positive because a $G2R_{1}$
event occurs, which makes $\alpha_{3}(t)$ larger. Then (\ref{type3di})
applies, where $x_{3,i}^{^{\prime}}(\tau_{k}^{-})=0$. Second, it is due to a
change of value in either $h_{1}(\tau_{k})$ or $\alpha_{1}(\tau_{k})$ or
$\beta_{3}(\tau_{k})$, which are all exogenous events.The state derivative is
the same as in (\ref{8b}).

This completes the derivation of all state and event time derivatives required
to evaluate the sample performance derivative in (\ref{dLi}). Using the
definition of $L_{n,m}(\theta)$ in (\ref{Lnm}), note that we can decompose
(\ref{dLi}) into its NEPs and evaluate the derivatives $dL_{n,m}%
(\theta)/d\theta_{i}$ as shown next.

\subsection{Cost Derivatives}

By virtue of (\ref{xprime(t)}), $x_{n,i}^{\prime}(t)$ is piecewise constant
during a NEP and its value changes only at an event point $t_{n,m}^{j}$,
$j=1,...,J_{n,m}$. Therefore, we have
\begin{align}
\frac{dL_{n,m}(\theta)}{d\theta_{i}} &  =\int\nolimits_{\xi_{n,m}(\theta
)}^{\eta_{n,m}(\theta)}x_{n,i}^{\prime}(t)dt\nonumber\\
&  =x_{n,i}^{\prime}((\xi_{n,m})^{+})(t_{n,m}^{1}-\xi_{n,m})+x_{n,i}^{\prime
}((t_{n,m}^{J_{n,m}})^{+})\nonumber\\
&  \cdot(\eta_{n,m}-t_{n,m}^{J_{n,m}})+\sum\limits_{j=2}^{J_{n,m}}%
x_{n,i}^{\prime}((t_{n,m}^{j})^{+})(t_{n,m}^{j}-t_{n,m}^{j-1})\label{costDrtv}%
\end{align}
Clearly, the state derivative at each event point is determined by
(\ref{jumps}) which in turn depends on the event type at $t_{n,m}^{j}$,
$j=1,...,J_{n,m}$ and is given by the corresponding expression in
(\ref{type1}) through (\ref{8b}). An explicit closed-form expression of
$dL_{n,m}(\theta)/d\theta_{i}$ may be obtained in this manner but becomes
complicated. A simple algorithm that updates $\frac{dL_{n,m}(\theta)}%
{d\theta_{i}}$ after every observed event is simple to implement. More
importantly, note that this IPA derivative depends on: $(i)$ the number of
events in each NEP $J_{n,m}$, $(ii)$ the number of total $G2R_{n}$ events
$\zeta_{n,k}$, $(iii)$ the number of total $R2G_{n}$ events $\rho_{n,k}$,
$(iv)$ the event times $\xi_{n,m}$, $\eta_{n,m}$ and $t_{n,m}^{j}$, and $(v)$
the arrival and departure rates $\alpha_{n}(\tau_{k})$, $\beta_{n}(\tau_{k})$
at an event time \emph{only}. The quantities in $(i)-(iv)$ are easily observed
through counters and timers. The rates in $(v)$ may be obtained through simple
estimators, emphasizing that they are only needed at a finite number of
observed event times.

\section{Simulation Results}

We describe how the IPA estimator derived for the SFM can be used to determine
optimal light cycles for two intersections modeled as a DES. We apply the IPA
estimator using actual data from an observed sample path of this DES (in this
case, by simulating as a pure DES).

We assume cars arrive according to a Poisson process with rate $\bar{\alpha
}_{n}$, $n=1,2,4$ (as already emphasized, our results are independent of this
distribution.). We also assume cars depart at a rate $h_{n}(t)$ which we fix
to be a constant $H_{n}$ when the road is not empty. We also constrain $\theta_{i},i=1,...,4$, to
take values in $[\theta_{\min},\theta_{\max}]$.

For the simulated DES model, we use a brute-force (BF) method to find an
optimal $\theta_{BF}^{\ast}$: we discretize all real values of $\theta_{i}$
and for $\theta_{i}, i=1,...,4$ combinations we run $10$ sample paths
to obtain the average total cost. The value of $\theta_{BF}^{\ast}$ is the one
generating the least average cost, to be compared to $\theta_{IPA}^{\ast}$,
the IPA-based method. In our simulations, we estimate $\alpha_{n}(\tau_{k})$ through ${N_{a}}/{t_{w}}$ by
counting car arrivals $N_{a}$ over a time window $t_{w}$ before or after
$\xi_{n,m}$; $\beta_{n}(\tau_{k})$ is similarly estimated.

\begin{figure}[tbh]
\centering
\includegraphics[scale = 0.4]{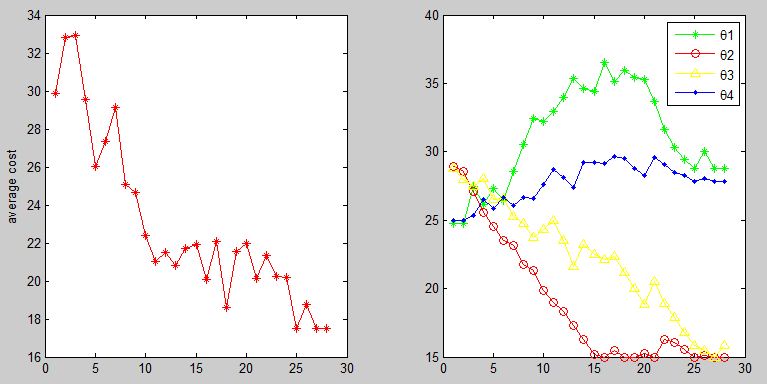}\caption{IPA-based TLC result I }%
\label{Trajectory}%
\end{figure}

In the results reported here, we set $\bar{\alpha}_{n}=1/4,n=1,2,4$,
$H_{n}=1,n=1,2,3,4$,  $\theta_{\min}=15sec,$ $\theta_{\max
}=40sec$ and the sample length $T=1000sec$. Fig. \ref{Trajectory} shows the trajectories of $J$ and
$\theta$ using the IPA-based method where $w=[10,1,1,1]$ and initial $\theta_0=[25,30,30,25]$ . More results are
shown in Table 1. As we can see, $\theta_{IPA}^{\ast}$ is approaching the
optimal value obtained by the BF method. Notice that BF method becomes
impractical when there are more controlling parameters, or when the range of
the parameter is large. However, the IPA method is still effective in such
situations. Moreover, we notice that the value of $(\theta_3+\theta_4)$ is similar to $(\theta_1+\theta_2)$. This indicates that the two intersections tend to have the sample traffic light switching cycle to balance traffic flows.

\begin{table}[tbh]
\caption{IPA vs BF method result I}%
\label{table1}	
\begin{center}%
\begin{tabular}
[c]{|c|c|c||c|c|}\hline
\multirow{2}{*}{w} & \multicolumn{2}{|c||}{BF} & \multicolumn{2}{|c|}{IPA}\\\cline{2-5}
 & $\theta^{*}$ & $J^{*}$ & $\theta^{*}$ & $J^{*}$\\\hline
{[1,1,1,1]}  & [15,15,15,15] & 5.4 & [15,15,15,15] & 5.4\\\hline
{[10,1,1,1]}  & [27,15,15,29] & 16.6 & [28.8,15,15,27.8] & 17.5\\\hline
{[1,5,5,1]}  & [15,23,17,21] & 12.6 & [15.1,18.6,15.6,18.5] & 13.2\\\hline
{[5,1,1,10]}  & [25,15,15,25] & 22.0 & [22.1,15,15,22.9] & 22.5\\\hline
{[1,10,1,1]}  & [15,29,15,29] & 16.3 & [15, 31.2,18.1,26.6] & 17.2\\\hline
\end{tabular}
\end{center}
\end{table}

Based on this observation, we also do simulations by setting $\theta_{1}+\theta_{2}=T_{1}$ and $\theta_{3}+\theta_{4}=T_{2}$, which
indicates that we set the \textquotedblleft GREEN plus RED\textquotedblright \ cycle to be fixed $T_{1}$ for each intersection. With this constraint, we only need to find optimal $\theta_{1}^{\ast}$ and $\theta_{3}^{\ast}$, since $\theta_{2}^{\ast}=T_{1}-\theta_{1}^{\ast}$ and $\theta_{4}^{\ast}=T_{2}-\theta_{3}^{\ast}$. We first let $T_{1}%
=T_{2}$, which restricts the two intersections to have the same traffic light switching cycle. Table 2 shows the simulation results. $T_1$ and $T_2$ are set to be the value obtained from Table 1. For example, when $w=[10,1,1,1]$, $\theta_1+\theta_2=42$ and $\theta_3+\theta_4=44$ in Table 1. We then set $T_1=T_2=44$ in Table 2, and restrict $\theta_{min}=15$. Comparing the results in Table 2 with Table 1, we find that it supports the results where we allow independent $\theta_i, i=1,2,...,4$.

\begin{figure}[tbh]
\centering
\includegraphics[scale = 0.4]{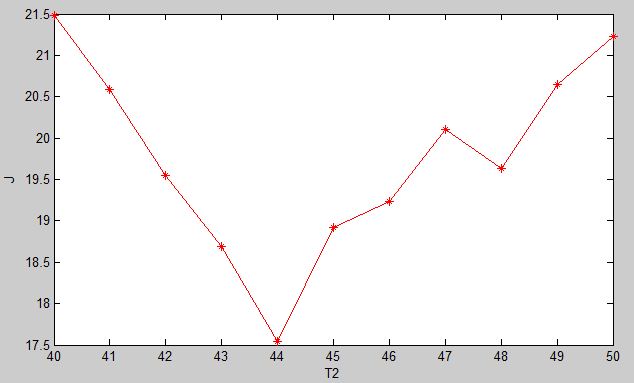}\caption{$J_{IPA}^{*}$ at different $T_2$}%
\label{T2}%
\end{figure}

In Fig. \ref{T2}, we set $w=[1,10,1,1]$ and change $T_2$ to obtain $J_{IPA}^{*}$ while keeping $T_1= 44$. As we can see, the minimum $J_{IPA}^{*}$ is achieved when $T_1=T_2$, which also matches the observations under independent $\theta_i$.

\begin{table}[tbh]
\caption{IPA vs BF method result II}%
\label{table2}
\begin{center}%
\begin{tabular}
[c]{|c|c|c|c||c|c|}\hline
\multirow{2}{*}{w} & \multirow{2}{*}{ $[T_1,T_2]$} & \multicolumn{2}{|c||}{BF}
& \multicolumn{2}{|c|}{IPA}\\\cline{3-6}
&  & $[\theta_{1}^{*},\theta_{3}^{*}]$ & $J^{*}$ & $[\theta_{1}^{*},\theta
_{3}^{*}]$ & $J^{*}$\\\hline
{[1,1,1,1]} & [30,30] & [15,15] & 5.4 & [15,15] & 5.4\\\hline
{[10,1,1,1]} & [44,44] & [31,20] & 16.2 & [30.4,21.6] & 17.6\\\hline
{[1,5,5,1]} & [39,39] & [15,16] & 12.2 & [15,15.6] & 14.1\\\hline
{[5,1,1,10]} & [40,40] & [25,15] & 24.3 & [23.1.6,12.5] & 24.3\\\hline
{[1,10,1,1]} & [44,44] & [15,16] & 17.5 & [15,15.2] & 17.6\\\hline
\end{tabular}
\end{center}
\end{table}

\begin{figure}[tbh]
\centering
\includegraphics[scale = 0.4]{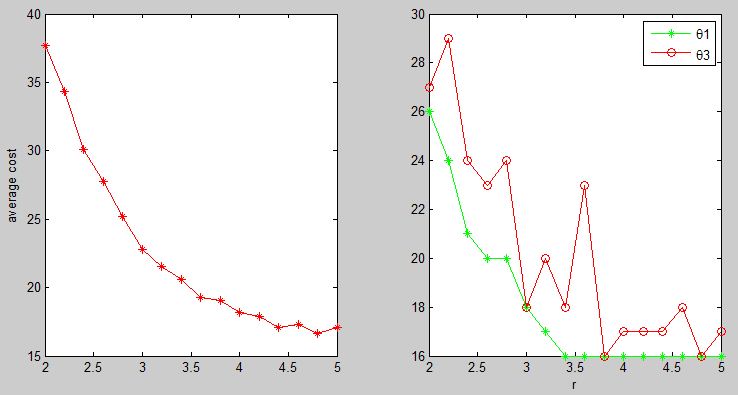}\caption{$J_{IPA}^{*}$ and $\theta_{IPA}^{*}$ at different $r$}%
\label{arrivalRate}%
\end{figure}

We are also interested to see the optimal control parameters under different traffic intensities. We set $w=[1,10,1,1]$, $T_1=T_2=44$ and $\alpha = [1/r,1/4,1/4]$, i.e., we operate under different arrival rate of queue 1. Fig. \ref{arrivalRate} shows the optimal cost and optimal $\theta_1$ and $\theta_3$ while $r$ varies. It is clear to see that $\theta_{1}^{*}$ increases as $r$ decreases. This indicates more GREEN light duration is assigned to queue $1$ as more cars are accumulated in queue 1 because of the fast arrival rate. $\theta_3^{*}$ also increases because more cars flow into queue 3. 

It must be pointed out that the BF method does not provide a \textquotedblleft true\textquotedblright\ optimal, since the DES model of the traffic system is as much an approximation as the SFM based on which IPA operates. Thus, the comparative results should be interpreted accordingly. 

\section{Conclusions and Future work}

We have developed a SFM for a traffic light control problem with two
coupledintersections, based on which we derive an IPA\ gradient estimator of a
cost metric with respect to the controllable green and red cycle lengths. The
estimators are used to iteratively adjust light cycle lengths to improve
performance and, under proper conditions, obtain optimal values which adapt to
changing traffic conditions. The analysis in the paper can be readily extended
to $N$ intersections in tandem. Future work will extend our method to solving
the TLC problem over multiple junctions without assumption $(iii)$, i.e.,
allowing a finite car capacity between intersections to cause blocking effects.


\bibliographystyle{plain}
\bibliography{yf}

\begin{thebibliography}{22}
\providecommand{\natexlab}[1]{#1}
\providecommand{\url}[1]{\texttt{#1}}
\providecommand{\urlprefix}{URL }
\expandafter\ifx\csname urlstyle\endcsname\relax
  \providecommand{\doi}[1]{doi:\discretionary{}{}{}#1}\else
  \providecommand{\doi}{doi:\discretionary{}{}{}\begingroup
  \urlstyle{rm}\Url}\fi

\bibitem[{Alvarez and Poznyak(2010)}]{Alvarez10}
Alvarez, I. and Poznyak, A. (2010).
\newblock Game theory applied to urban traffic control problem.
\newblock \emph{Intertional Conference on Control, Automation and Systems},
  2164--2169.

\bibitem[{Cassandras and Lafortune(2008)}]{Cassandras08}
Cassandras, C.G. and Lafortune, S. (2008).
\newblock \emph{Introduction to Discrete Event Systems, 2nd ed.}
\newblock Springer.

\bibitem[{Cassandras and Lygeros(2006)}]{Cassandras06}
Cassandras, C.G. and Lygeros, J. (2006).
\newblock \emph{Stochastic Hybrid Systems}.
\newblock Taylor and Francis.

\bibitem[{Cassandras et~al.(2002)Cassandras, Wardi, Melamed, Sun, and
  Panayiotou}]{Cassandras02}
Cassandras, C.G., Wardi, Y., Melamed, B., Sun, G., and Panayiotou, C.G. (2002).
\newblock Perturbation analysis for on-line control and optimization of
  stochastic fluid models.
\newblock \emph{IEEE Trans. Automat. Control}, 47(8), 1234--1248.

\bibitem[{Cassandras et~al.(2010)Cassandras, Wardi, Panayiotou, and
  Yao}]{Cassandras10}
Cassandras, C.G., Wardi, Y., Panayiotou, C.G., and Yao, C. (2010).
\newblock Perturbation analysis and optimization of stochastic hybrid systems.
\newblock \emph{Europ. J. of Control}, 16(6), 642--664.

\bibitem[{Choi et~al.(2002)Choi, Yoon, Kim, Chung, and Lee}]{Choi02}
Choi, W., Yoon, H., Kim, K., Chung, I., and Lee, S. (2002).
\newblock A traffic light controlling flc considering the traffic congestion.
\newblock \emph{AFSS 2002, International Conference on Fuzzy Systems}, 69--75.

\bibitem[{DeSchutter(1999)}]{Schutter99}
DeSchutter, B. (1999).
\newblock Optimal traffic light control for a single intersection.
\newblock \emph{Proceedings of the IEEE American Control Conference}, 3,
  2195--2199.

\bibitem[{Dujardin et~al.(2011)Dujardin, Boillot, Vanderpooten, and
  Vinant}]{Dujardin11}
Dujardin, Y., Boillot, F., Vanderpooten, D., and Vinant, P. (2011).
\newblock Multiobjective and multimodal adaptive traffic light control on
  single junctions.
\newblock \emph{14th IEEE Conference on Intelligent Transportation Systems},
  1361--1368.

\bibitem[{Findler and Stapp(1992)}]{Findler92}
Findler, N. and Stapp, J. (1992).
\newblock A distributed approach to optimized control of street traffic
  signals.
\newblock \emph{Journal of Transportation Engineering}, 118.

\bibitem[{Geng and Cassandras(2012)}]{GengCDC12}
Geng, Y. and Cassandras, C.G. (2012).
\newblock Traffic light control using infinitesimal perturbation analysis.
\newblock Technical report.

\bibitem[{Head et~al.(1996)Head, Ciarallo, and an~V.~Kaduwela}]{Head96}
Head, L., Ciarallo, F., and an~V.~Kaduwela, D.L. (1996).
\newblock A perturbation analysis approach to traffic signal optimization.
\newblock \emph{INFORMS National Meeting}.

\bibitem[{Murat and Gedizlioglu(2005)}]{Murat05}
Murat, Y.S. and Gedizlioglu, E. (2005).
\newblock A fuzzy logic multi-phased signal control model for isolated
  junctions.
\newblock \emph{Transportation Research Part C}, 18, 19--36.

\bibitem[{Panayiotou et~al.(2005)Panayiotou, Howell, and Fu}]{Panayiotou05}
Panayiotou, C.G., Howell, W.C., and Fu, M. (2005).
\newblock Online traffic light control through gradient estimation using
  stochastic fluid models.
\newblock \emph{Proceedings of the IFAC 16th Triennial World Congress}.

\bibitem[{Prashanth and Bhatnagar(2011)}]{Prashanth11}
Prashanth, L. and Bhatnagar, S. (2011).
\newblock Reinforcement learning with average cost for adaptive control of
  traffic lights at intersections.
\newblock \emph{14th IEEE Conference on Intelligent Transportation Systems},
  1640--1645.

\bibitem[{Taale et~al.(1998)Taale, Back, Preub, Eiben, Graaf, and
  Schippers}]{Taale98}
Taale, H., Back, T., Preub, M., Eiben, A., Graaf, J., and Schippers, C. (1998).
\newblock Optimizing traffic light controllers by means of evolutionary
  algorithms.
\newblock \emph{EUFIT}.

\bibitem[{Thorpe(1997)}]{Thorpe97}
Thorpe, T. (1997).
\newblock \emph{Vehicle traffic light control using sarsa}.
\newblock Master thesis, Dept. of Comp. Sci., Colorado State Univ.

\bibitem[{Wardi et~al.(2010)Wardi, Adams, and Melamed}]{Wardi10}
Wardi, Y., Adams, R., and Melamed, B. (2010).
\newblock A unified approach to infinitesimal perturbation analysis in
  stochastic flow models: the single-stage case.
\newblock \emph{IEEE Trans. Automat. Control}, 55(1), 89--103.

\bibitem[{Wiering et~al.(2004)Wiering, Veenen, Vreeken, and
  Koopman}]{Wiering04}
Wiering, M., Veenen, J., Vreeken, J., and Koopman, A. (2004).
\newblock Intelligent traffic light control.
\newblock \emph{Technical Report UU-CS-2004}.

\bibitem[{Yao and Cassandras(2011{\natexlab{a}})}]{Yao11}
Yao, C. and Cassandras, C.G. (2011{\natexlab{a}}).
\newblock Resource contention games in multiclass stochastic flow models.
\newblock \emph{Nonlinear Analysis: Hybrid Systems}, 5(2), 301--319.

\bibitem[{Yao and Cassandras(2011{\natexlab{b}})}]{YaoCgc11}
Yao, C. and Cassandras, C. (2011{\natexlab{b}}).
\newblock Perturbation analysis of stochastic hybrid systems and applications
  to resource contention games.
\newblock \emph{Frontiers of Electrical and Electronic Engineering in China},
  6(3), 453--467.

\bibitem[{Yu and Recker(2006)}]{Yu06}
Yu, X. and Recker, W. (2006).
\newblock Stochastic adaptive control model for traffic signal systems.
\newblock \emph{Transportation Research Part C: Emerging Technology}, 14(4),
  263--282.

\bibitem[{Zhao and Chen(2003)}]{Zhao03}
Zhao, X. and Chen, Y. (2003).
\newblock Traffic light control method for a single intersection based on
  hybrid systems.
\newblock \emph{Proc. of the IEEE Intelligent Transp. Systems}, 1105--1109.

\end{thebibliography}








\end{document}